\documentclass[reprint,amsmath,amssymb,superscriptaddress,aps,pre]{revtex4-2}
\usepackage{geometry}
\usepackage{comment}
\usepackage{enumerate}
\usepackage{amsmath,nccmath}
\usepackage{xcolor}
\usepackage{mathrsfs}
\usepackage{graphicx}% Include figure files
\usepackage{dcolumn}% Align table columns on decimal point

\usepackage{natbib}
\usepackage{bm}% bold math

\usepackage[mathlines]{lineno}% Enable numbering of text and display math
\usepackage{xcolor}
\usepackage{color}
\usepackage[colorlinks=true,linktoc=page,citecolor=blue,linkcolor=blue,]{hyperref}
\newcommand{\bb}{{\bm b}}
\newcommand{\bu}{{\bm u}}
\newcommand{\bx}{{\bm x}}
\newcommand{\br}{{\bm r}}
\begin{document}

\title{Energy transfer in simple and active binary fluid turbulence {\bf {-}} a false friend of incompressible MHD turbulence } 
\author{Nandita Pan}
\author{Supratik Banerjee}
\email{sbanerjee@iitk.ac.in}
\affiliation{
 Department of Physics, Indian Institute of Technology Kanpur, Uttar Pradesh, 208016 India
}%

\date{\today}

\begin{abstract}
 Inertial range energy transfer in three dimensional fully developed binary fluid turbulence is studied under the assumption of statistical homogeneity. Using two point statistics, exact relations corresponding to the energy cascade are derived (i) in terms of two-point increments and (ii) two-point correlators. Despite having some apparent resemblances, the exact relation in binary fluid turbulence is found to be different from that of the incompressible MHD turbulence (Politano and Pouquet, GRL, 1998). 
  Besides the usual direct cascade of energy, under certain situations, an inverse cascade of energy is also speculated depending upon the strength of the activity parameter and the interplay between the two-point increments of the fluid velocity and the composition gradient fields. An alternative form of the exact relation is also derived in terms of the `upsilon' variables and a subsequent phenomenology is also proposed predicting a $k^{-{3}/{2}}$ law for the turbulent energy spectrum. 
 \end{abstract}

\maketitle
\section{Introduction}
Turbulence is a ubiquitous phenomenon in classical fluids, typically characterised by its multiscale vortex-like structures and a universal cascade of inviscid invariants (energy, helicity etc.) within the inertial range. For homogeneous and isotropic turbulence of an incompressible fluid, the energy cascade leads to a $k^{-5/3}$ spectra \citep{kolmogorov1941local} whereas, for isotropic MHD turbulence, a $k^{-3/2}$ power spectrum is predicted \citep{Iroshnikov1964turbulence,Kraichnan1965Inertial}.  In real space, the universality of turbulence can be understood in terms of exact scaling relations which express the average flux rate $(\varepsilon)$ of a cascading invariant in terms of the statistical moments of two-point fluctuations of the relevant field variables. Under the assumption of homogeneity and isotropy, a few algebraic exact relations are obtained for incompressible hydrodynamic (HD) and magnetohydrodynamic (MHD) turbulence \citep{kolmogorov1941degeneration, politano1998karman,gomez2000exact}. However, for homogeneous (and not necessarily isotropic) turbulent flows, differential exact laws have been derived with the generic expression 
\begin{equation}
 \bm{\nabla}_{\bm r} \cdot\bm{\mathcal{F}} + \bm{\mathcal S} = - 4 \varepsilon, \label{eq1}  
\end{equation} 
\noindent where $\bm {\mathcal{F}}$ denotes the flux term involving the statistical moments of the two-point increments, $\bm{\mathcal S}$ denotes the source term and $\bm r$ represents the two-point separation. For incompressible HD and MHD turbulence, it has been found \citep{monin_jaglom_1975, politano1998dynamical} that $\bm{\mathcal S} =0$, and 
\begin{align}
  \bm{\mathcal{F}_{HD}} &=  \langle \left(\delta \bm{u}\right)^2 \delta \bm{u}\rangle, \\ 
 \bm{\mathcal{F}_{MHD}} &=  \langle \left[(\delta \bm{u})^2 + {\color{black}(\delta \bm{b})^2} \right] \delta \bm{u} - 2 (\delta \bm{u} \cdot {\color{black}\delta \bm{b}}) {\color{black}\delta \bm{b}} \rangle, \label{Incompressible_MHD}
\end{align} 
\noindent where $\bu$ and $\bb$ denote the fluid velocity and the magnetic field (normalized to a velocity) respectively, and $\delta {\bm a} = {\bm a} ( \bx + \br ) - {\bm a} (\bx)$ represents the two-point increment for the variable ${\bm a}$. The corresponding algebraic exact relations can finally be obtained just by integrating the divergence term under spherical symmetry \cite{antonia1997analogy,banerjee2014compressible}. In addition to the abovesaid simple turbulent flows, Eq.~\eqref{eq1} represents the turbulent energy transfer for a broader range of fluids including incompressible Hall-MHD plasma and compressible fluids and plasmas where the source term is non-vanishing and the derivation of isotropic algebraic forms is not straightforward \citep{galtier2011exact, banerjee2013exact, banerjee2014kolmogorov,andres2019statistics}. However, algebraic exact relations for a large number of homogeneous turbulent flows can be derived following an alternative formulation proposed recently, for the first time, for incompressible HD and MHD (including Hall MHD) turbulence \citep{banerjee2016alternative, banerjee2016chiral} and then generalized to more complicated systems e.g. rotating fluid, compressible fluids and plasmas \citep{banerjee2018energy, banerjee2020scale} and even ferrofluids \citep{mouraya2019determination}. Besides the vector fields (velocity and magnetic fields), often it is interesting to study the behaviour of a scalar field $\phi $ in a turbulent flow. In general, the scalar field is advected by the velocity field satisfying usually the advection-diffusion equation. Furthermore, it can be passive ($e.g.,$ dust particles in a turbulent flow) or active ($e.g.,$ stratified flows) depending on whether it provides feedback to the momentum equation \citep{verma2019Energy}.
{\color{black}Incompressible flows with a passive scalar} possess two independent quadratic inviscid invariants, the kinetic energy $\frac{1}{2}\langle\bu^2\rangle$ and {\color{black} the scalar energy}  $\frac{1}{2}\langle\phi^2\rangle$. While the first conservation leads to an exact relation similar to incompressible HD turbulence, the second one leads to a separate differential exact relation with $\bm{\mathcal F} = \langle(\delta \phi)^2\delta \bm{u} \rangle$ and $\bm{\mathcal S} = 0$ \citep{Yaglom1949}. 

{\color{black} For active scalars, the kinetic energy is no longer conserved}. The simplest active scalar flow is represented by a stably stratified fluid where the active scalar $ \phi$ denotes the density perturbation and provides a feedback force density proportional to $\phi\hat{z}$, with $\hat{z}$ being the direction of stratification. For such flows, $\frac{1}{2}\langle\phi^2\rangle$ remains to be an inviscid invariant. However, instead of the kinetic energy, the total energy $\frac{1}{2}\langle u^2+ \phi^2/N^2\rangle$ is found to be an inviscid invariant and the corresponding source-less differential exact relation is given by $\bm{\mathcal{F}} = \langle[(\delta \bm{u})^2 + (\delta \phi /N)^2]\delta \bm{u} \rangle$ \citep{augier2012kolmogorov}, where $N$ is the Brunt-V\"{a}is\"{a}l\"{a} frequency. 

A more complex feedback force arises if we consider a system of binary fluid mixtures. Binary fluid is a two-component system ranging from a mixture of two simple fluids $e.g.,$ oil and water, to complex systems such as ‘active' binary fluids \citep{hohenberg1977critical,Ruiz1981turbulence, cates2018theories,tiribocchi2015active} {\footnote{the word ‘active' is used here with a different meaning from that of an ‘active' scalar and will be explained {\color{black}later}}}. For such systems, the active scalar $\phi$ represents the local molecular composition of the binary mixture \citep{cates2018theories}. {\color{black}Simple binary fluids contain microstructures like droplets or domains of one fluid into the sea of the other thus producing interfaces or domain walls.} Sharp variation in $\phi$ across the interfaces generates diffusion currents which, in turn, drive the velocity field by exerting feedback stress proportional to $\kappa(\bm{\nabla}\phi \otimes \bm{\nabla}\phi)$ \citep{kendon2001inertial,cates2018theories}, where $\kappa$ is a positive constant. However feedback stress does not necessarily take the same form if one of the binary-fluid components is `active' \textit{i.e.} the constituent particles are equipped with intrinsic mechanisms ($e.g.,$ self-propulsion, body deformation \citep{wolgemuth2008collective, ohta2017dynamics} etc.,) of transmuting ambient free energy into directed motion thereby breaking the time reversal symmetry (TRS) at the scale of each constituent \citep{fodor2016far}. Nevertheless,
if the active particles are either spherically symmetric or possess a weak orientational order ($e.g.,$ artificial active colloides, dilute bacterial suspensions etc.,) the feedback stress due to the activity of the particles can be written as $\eta(\bm{\nabla}\phi\otimes\bm{\nabla}\phi)$ where $\eta$ is the activity parameter \citep{cates2013active}. The total feedback stress is finally obtained by adding the active stress to the stress due to the interfacial tension, and is given by $(\kappa+\eta)(\bm{\nabla}\phi\otimes\bm{\nabla}\phi)$ \citep{tiribocchi2015active, cates2018theories}. 

Although individual components of a binary fluid tend to phase separate {(\color{black} coarse-graning)} below a critical temperature ($T_c$) \citep{chaikincambridge}, it is often useful to have them in the form of steady emulsions  \textit{i.e.} a homogeneous phase-mixed state \citep{Bibette1999emulsions}. This can be achieved through the generation of turbulence which, in effect, prevents the phase separation owing to its enhanced mixing property {\color{black}\citep{pine1984turbulent,perlekar2014spinodal, pal2016binary, bhattacharjee2021activity}}. {\color{black}In a physical system, turbulence can be attained in two ways, either by large non-linear perturbations ($e.g.,$ MHD, stably stratified flows etc.,) or due to the growth of linear instability ($e.g.,$ Rayleigh-Ben\'{a}rd convection)}. Simple binary fluid and active fluid with extensile stress ($\eta>0$) belong to the former category \citep{Ruiz1981turbulence,tiribocchi2015active} whereas active fluid with contractile stress ($\eta<0$) having large activity ($i.e. \, \, \xi = (\kappa+\eta) < 0$) belongs to the latter one \citep{kirkpatrick2019driven, das2020transition}. Turbulence in an active binary fluid with $\xi<0$ finds its application in the study of bioconvective plums, accumulation of $Dinoflagellates$ in turbulent flows $etc.,$ \citep{pedley1992hydrodynamic}. Note that here, we are referring to high Reynolds number ($Re$) turbulence in active fluids which is different from the self-sustained pattern formation by dense active fluids at low $Re$ ($\sim 10^{-5}$) {\color{black}\citep{dombrowski2004self,wensink2012meso,dunkel2013minimal,james2018turbulence, linkmann2019phase, sanchez2012spontaneous,doostmohammadi2018active,alert2021active}}. The effect of turbulence on domain growth, energy spectra etc., have been explored both for simple and active binary fluids \citep{Ruiz1981turbulence, aronovitz1984turbulence, berti2005turbulence, perlekar2019kinetic, tiribocchi2015active, bhattacharjee2021activity}. However, till date, no exact relation has been derived for such systems. 

In this paper, using two-point statistics, we derive exact relations for the inertial range energy transfer in binary fluid turbulence (BFT). Under the assumption of statistical homogeneity, first we derive a differential exact relation as in Eq.~\eqref{eq1} in terms of the two-point fluctuations of ${\bf u}$ and ${\bm \nabla \phi}$. Then the same exact relation is also expressed in terms of (i) two-point correlators and (ii) newly defined upsilon variables. The derivation of the results are complemented by comparative statements on incompressible MHD turbulence which seems to have some interesting resemblances with our system \citep{Ruiz1981turbulence}.

The paper is organized as follows: in Sec. \ref{Basic Equations}, we describe the model and the basic equations of dynamics followed by the derivation of total energy conservation in the inviscid limit. Sec. \ref{Exact Relation} contains the detailed derivation of the exact relation both in terms of the two-point increments and two-point correlators. We then introduce a set of variables ($\bm{\Upsilon^{\pm}}$) and express the inviscid invariant and the corresponding exact relation in terms of those variables in Sec. \ref{upsilon variables}. Finally in Sec. \ref{Discussion}, we discuss our results and conclude.

\section{Model and governing equations}\label{Basic Equations}

%The dynamics of Phase separated or immiscible binary fluid is mainly governed by the CH-NS equations (Cahn-Hilliard (CH) equation governing evolution of $\phi(\bm{r},t)$ coupled to the reactive NS equation with feedback $\bm{\Sigma}$). Primary focus of our study is to investigate the nature of conserved energy transfer in real space for a fully-developed turbulent flow. This can be effectively done by the coupled dynamics of AD (a simplified CH equation) and reactive NS equation. AD-NS equation is mainly applicable to miscible binary fluids \citep{Ruiz1981turbulence,kirkpatrick2019driven,bhattacharjee2021activity}. However, the energy transfer due to phase separation in CH equation comes through dissipation only, which is anyway negligible within inertial range. Hence the conserved energy transfer for
 %both miscible and immicsible simple and active binary fluids 
%can be well-studied using AD-NS equation.

 A binary fluid, composed of two components $A$ and $B$, is usually defined in terms of their mean velocity field $\bm{u}(\bm{x},t) = (\rho_A \bm{u}_A + \rho_B \bm{u}_B)/(\rho_A+\rho_B)$ and the local molecular composition $\phi(\bm{x},t)= (\rho_A-\rho_B)/(\rho_A+\rho_B)$, where ($\rho_A$, $\rho_B$) and ($\bm{u}_A$, $\bm{u}_B$) are the densities and the velocities of the components $A$ and $B$ respectively \citep{cates2018theories}. Here we are considering systems where the velocity fluctuations are much less than the sound speed, thus justifying the assumption of incompressiblity for the resultant fluid \citep{Ruiz1981turbulence}. The corresponding continuity and momentum evolution equations are given by 
\begin{align}
 \bm{\nabla} \cdot \bm{u} &= 0, \label{incompressibilty}\\
\partial_{t} \bm{u} + (\bm{u}\cdot\bm{\nabla}) \bm{u} &= -\bm{\nabla} p  + \bm{\nabla}\cdot\bm{\Sigma} + \nu \nabla^{2} \bm{u} +\bm{f}, \label{E_momentum}
\end{align} 
where $p$ is the fluid pressure, $\bm{\Sigma}$ the feedback stress tensor due to the active scalar $\phi$, $\nu$ the kinematic viscosity and finally $\bm{f}$ represents a large scale forcing. For a general phase-separating binary fluid mixture, $\phi$ satisfies Cahn-Hilliard equation given by
\begin{align}
 \partial_{t} \phi + (\bm{u}\cdot\bm{\nabla}) \phi &= \mathcal{M}\nabla^2 \mu, \label{cahn1}\\
\mu = \frac{\delta \mathcal{F}[\phi]}{\delta \phi} &= a\phi + b\phi^3 - \kappa \nabla^2 \phi,\label{mu}
\end{align}
where $\mathcal{M}$ is the mobility coefficient, $\mu$ the chemical potential (or exchange potential) and  $\mathcal{F} = \int \left[ \frac{a}{2} \phi^2 + \frac{{\color{black}b}}{4}\phi^4  + \kappa (\bm{\nabla} \phi)^2 \right]d\tau$ represents a free energy functional, where $a<0$ while $b$ and $\kappa$ are positive constants and $\tau$ represents the volume \citep{cahn1968Spinodal,hohenberg1977critical,cates2018theories}. For a phase-mixed binary fluid, $a>0$ and $\mathcal{F} = \int \left[ \frac{a}{2} \phi^2 + \kappa (\bm{\nabla} \phi)^2 \right]d\tau$ and the Eq. (\ref{cahn1}) simply becomes
\begin{align}
      \partial_{t} \phi + (\bm{u}\cdot\bm{\nabla}) \phi &= \mathcal{M}\nabla^2 (a\phi-\kappa \nabla^2 \phi),\label{hyperviscosity}
\end{align}
In addition to the force equation \eqref{E_momentum}, one can also force Eq.~\eqref{cahn1} or Eq.~\eqref{hyperviscosity} by an appropriate large scale forcing $g_\phi$ \citep{bhattacharjee2021activity}. 
 The corresponding evolution equation of $\phi(\bm{x},t)$ is therefore given by 
\begin{align}
\partial_{t} \phi + (\bm{u}\cdot\bm{\nabla}) \phi &= \mathcal{M}\nabla^2\mu + g_\phi.\label{E_phi1}
 \end{align}
The final step is to express the feedback stress tensor $\bm{\Sigma}$ in terms of the other field variables.
Both for phase-separating and phase-mixed binary fluids, local inhomogeneity in $\phi$ leads to a feedback force density $\bm{\nabla}\cdot\bm{\Sigma} = - \phi \bm{\nabla}\mu $ \citep{kendon2001inertial,cates2018theories}.
Using the expressions of $\mu$ and following a few steps of straightforward algebra, the feedback stress can be put in the form given below:
\begin{align}
    \bm{\Sigma} &= - \left[\kappa(\bm{\nabla} \phi \otimes \bm{\nabla} \phi) + \Lambda \mathbb{I} \right]\label{Passive_stress},
\end{align}
where $\sqrt{\kappa}$ is dimensionally homogeneous to the kinematic viscosity \citep{Ruiz1981turbulence} and 
\begin{equation}
    \Lambda = \left(\phi\frac{d\Gamma}{d\phi}-\Gamma-\kappa\phi\nabla^2\phi-\kappa\frac{(\bm{\nabla} \phi )^{2}}{2}\right),
\end{equation}
with $\Gamma$ being the $\phi$ dependent part of the free energy {\color{black}density}. Evidently, the term $\Lambda \mathbb{I}$ is proportional to the unit tensor and can therefore be absorbed under a gradient in the force equation \eqref{E_momentum} which can now be expressed as:
\begin{align}
\partial_{t} \bm{u} + (\bm{u}\cdot\bm{\nabla}) \bm{u} = &-\bm{\nabla} P^*  -\kappa \bm{\nabla}\cdot(\bm{\nabla} \phi \otimes \bm{\nabla} \phi)\nonumber\\
&+\nu\nabla^2\bm{u}+\bm{f}, \label{E_momentum2}
\end{align}
where $P^*=p+ \Lambda$ can be understood as an effective pressure.  

When one of the binary fluid components is active, the additional stress arises due to the intrinsic swimming mechanism of active particles \citep{marchetti2013hydrodynamics}. {\color{black}Self-motility \textit{i.e.,} the ability of active particles to self-propel} produces a dipolar force field which yields the additional stress for active fluids.
If the particles are either spherically symmetric or having negligible orientational order, the active stress becomes proportional to $\eta \left(\bm{\nabla} \phi\otimes\bm{\nabla} \phi\right)$, where $\eta$ can be both positive or negative \citep{bhattacharjee2021activity}. Eq. \eqref{E_momentum2} now becomes:
\begin{align}
\partial_{t} \bm{u} + (\bm{u}\cdot\bm{\nabla}) \bm{u} = &-\bm{\nabla} P^*  -\xi \bm{\nabla}\cdot(\bm{\nabla} \phi \otimes \bm{\nabla} \phi)\nonumber\\
&+\nu\nabla^2\bm{u}+\bm{f}, \label{E_momentum3}
\end{align}
where $\xi=\kappa+\eta$. Note that, unlike the feedback stress for simple binary fluids, the additional ‘active’ stress is not derived from any free-energy functional \citep{cates2018theories}.  
For the sake of algebraic manipulation, it is convenient to consider $\bm{\nabla}\phi$ as a field variable rather than $\phi$ itself. The respective evolution equation is given by
\begin{align}
\partial_{t} (\bm{\nabla}\phi) + \bm{\nabla} (\bm{u}\cdot\bm{\nabla} \phi) &= \mathcal{M}\nabla^2\left(\bm{\nabla}\mu\right) + \bm{\nabla}g_\phi.    \label{E_phi2}
\end{align}
In the current study, we are interested in the turbulent energy transfer which necessarily involves the fluctuations with respect to the mean fields. By choosing an appropriate Galilean transformation, one can eliminate the mean velocity field. However, the mean composition gradient field can not be eliminated by such transformations. It is then useful to decompose $\bm{\nabla}\phi$ as
\begin{align}
\bm{\nabla}\phi = S\hat{z} + \bm{\nabla}\psi = S\hat{z} + \bm{q}, \label{phi_decompose}
\end{align}
where $S\hat{z}$ denotes the mean composition gradient field and $\bm{\nabla}\psi$ or $\bm{q}$ denotes the corresponding fluctuating field. The corresponding evolution equations of $\bm{u}$ and $\bm{q}$ will then be given by
\begin{align}
&\partial_{t} \bm{u} + (\bm{u}\cdot\bm{\nabla}) \bm{u} = -\bm{\nabla} P - \xi (S \hat{z} + \bm{q}) \bm{\nabla}\cdot \bm{q} + \nu\nabla^2\bm{u}+\bm{f},  \label{E_momentum4}\\
&\partial_{t}\bm{q} + \bm{\nabla}(Su_z+\bm{u}\cdot \bm{q})   = \mathcal{M}\nabla^2\left(\bm{\nabla}\mu\right) + \bm{g} \label{E_phi3},
\end{align}
where $\bm{\nabla}\cdot\bm{u} = 0$, $\bm{\nabla}\times\bm{q}=0$, $\bm{\nabla}g_\phi=\bm{g}$, $P=P^*+\xi(q^2/2+Sq_z)$ and $\bm{\nabla}\cdot[(S\hat{z}+\bm{q})\otimes(S\hat{z}+\bm{q})]=(S\hat{z}+\bm{q})\bm{\nabla}\cdot\bm{q}+\bm{\nabla}(q^2/2+Sq_z)$. 
For such a system, the total turbulent energy is composed of the kinetic and the `active' energy and can be written as
\begin{align}
    E = \int \frac{1}{2} \left( u^2 + \xi \bm{q}^2 \right) d\tau.\label{total_energy}
\end{align}
In the following, we show E is an inviscid invariant of the flow. For that we simply neglect the large scale forcing and small scale dissipation terms in Eqs. \eqref{E_momentum4} and \eqref{E_phi3}, which gives
\begin{align}
    &\partial_{t} \left(\frac{u^2}{2}\right)=\bm{u}\cdot\partial_{t} \bm{u} \nonumber\\ 
    &= - \bm{\nabla} \cdot \left(P+  \frac{u^2}{2}\right)\bm{u}
    - \xi (Su_z+ \bm{u} \cdot \bm{q}) \bm{\nabla}\cdot\bm{q},  \label{u2}\\
    &\partial_{t}\left(\frac{q^2}{2}\right)= \bm{q}\cdot\partial_{t} \bm{q} = -  \xi\bm{q}\cdot\bm{\nabla}\left(Su_z+ \bm{u} \cdot \bm{q} \right).  \label{q2}
\end{align}
Now combining Eqs. \eqref{total_energy}--\eqref{q2}, we obtain
\begin{align}
 & d_t E = \int \partial_{t} \left(\frac{u^2}{2} +\xi \frac{q^2}{2}\right)d \tau \nonumber\\
&=-\int \bm{\nabla} \cdot \left[P\bm{u}+  \frac{u^2}{2}\bm{u} +\xi ( Su_z+ \bm{u} \cdot \bm{q}) \bm{q} \right] d \tau.
\label{Conserved_Energy_Dev}
\end{align}
Finally using Gauss-divergence theorem with periodic or vanishing boundary conditions, one can show the total energy to be an inviscid invariant of the flow.

\section{Derivation of exact relation}\label{Exact Relation}
%A turbulent state is defined by the flux rate of an inviscid invariant of the flow. We have shown the conservation of total energy in previous section. 
Here we derive the two-point exact relation corresponding to the inertial range energy transfer in statistically homogeneous BFT.
Following \citep{banerjee2016alternative, banerjee2014compressible}, one can first define the two point correlator associated with the total energy (Eq. \eqref{total_energy}) as
\begin{equation}
\mathcal{R}_E =\mathcal{R}^{\prime}_E = \left \langle\frac{\bm{u}\cdot \bm{u}^\prime + \xi  \bm{q}\cdot \bm{q}^\prime }{2}  \right \rangle. \label{Correlator}
\end{equation}
where the unprimed and primed  quantities represent the corresponding field properties at point $\bm{x}$ and $\bm{x}^\prime = \bm{x+r}$ respectively. \par %For statistical homogeneity, one can have identity is true and will be useful for further manipulations \citep{frisch1995turbulence}-  \begin{equation}
%\bm{\nabla} \cdot \langle(\cdot)\rangle = - \bm{\nabla}_{\bm{r}}\cdot \langle(\cdot)\rangle = - \bm{\nabla}^{\prime} \cdot \langle(\cdot)\rangle. \label{Del}
%\end{equation} 
Now, we calculate the time evolution of the energy correlators. Similar to Eq. \eqref{E_momentum4}, we can also write the evolution equation for $\bm{u}^\prime$. Now combining $\bm{u}\cdot\partial_t\bm{u}^\prime$ and $\bm{u}^\prime\cdot\partial_t\bm{u}$, we obtain
\begin{align}
\partial_t &\left\langle \bm{u}\cdot \bm{u}^\prime \right\rangle \nonumber\\
=&\left\langle \bm{u}^\prime \cdot[-(\bm{u}\cdot\bm{\nabla}) \bm{u}  -  \bm{\nabla} P  -\xi(S\hat{z}+\bm{q} )
\bm{\nabla}\cdot\bm{q}] \right. \nonumber
\\
 &+ \left.\bm{u}\cdot [-(\bm{u}^\prime\cdot\bm{\nabla}^\prime) \bm{u}^\prime  -  \bm{\nabla}^\prime P^\prime -\xi(S\hat{z}+ \bm{q}^\prime ) \bm{\nabla}^\prime \cdot\bm{q}^\prime]\right \rangle\nonumber\\
&+ D_u +F_u\label{RE_u1} \\
 =& -\left\langle  \bm{\nabla}\cdot[(\bm{u}\cdot \bm{u}^\prime)\bm{u} + \xi (S u^\prime_z\bm{q}+(\bm{u}^\prime\cdot  \bm{q})\bm{q})]\right.
 \nonumber
\\
& +\left. \bm{\nabla}^\prime\cdot[( \bm{u}\cdot \bm{u}^\prime)\bm{u}^\prime +\xi(S u_z\bm{q}^\prime+ (\bm{u}\cdot\bm{q}^\prime)\bm{q}^\prime )]\right\rangle \nonumber\\
 & +D_u +F_u\label{RE_u2}\\
 =&-\bm{\nabla}_{\bm{r}}\cdot\left\langle  \bm{u}^\prime(\bm{u}\cdot\bm{u}^\prime)-\bm{u}(\bm{u}\cdot\bm{u}^\prime)+\xi Su_z\bm{q}^\prime-\xi Su_z^\prime\bm{q}\right.\nonumber\\
& +\left.(\bm{u}\cdot\bm{q}^\prime)\bm{q}^\prime-(\bm{u}^\prime\cdot\bm{q})\bm{q} \right\rangle + D_u + F_u,
 \label{RE_u}
\end{align}
where, $D_u = \langle\bm{u}^\prime\cdot\nu\nabla^2\bm{u}+ \bm{u}\cdot\nu\nabla^2\bm{u}^\prime\rangle$ and  $F_u = \langle\bm{u}^\prime\cdot\bm{f}+\bm{u}\cdot\bm{f}^\prime\rangle$ represent the effective dissipation and forcing contributions in $\partial_t \left\langle \bm{u}\cdot \bm{u}^\prime \right\rangle$. To obtain Eq.~\eqref{RE_u}, we also use the property of statistical homogeneity 
\begin{equation}
\bm{\nabla} \cdot \langle(\cdot)\rangle = - \bm{\nabla}_{\bm{r}}\cdot \langle(\cdot)\rangle = - \bm{\nabla}^{\prime} \cdot \langle(\cdot)\rangle, \label{Del}
\end{equation} 
and the following relations:
\begin{align*}
   &(i) \,\bm{u}^\prime\cdot \bm{q} (\bm{\nabla} \cdot \bm{q}) = \bm{\nabla}\cdot [ (\bm{u}^\prime \cdot \bm{q})\bm{q}] - \bm{\nabla}\cdot\left(\bm{u}^\prime \frac{q^2}{2}\right),\\
   &(ii)\, \langle \bm{u}\cdot \bm{\nabla}^\prime P^\prime \rangle = -\langle P^\prime (\bm{\nabla}\cdot \bm{u})\rangle=0,\\
   &(iii)\, \left\langle\bm{\nabla}\cdot\left(\bm{u}^\prime \frac{q^2}{2}\right)\right\rangle = - \left\langle \frac{q^2}{2}\left(\bm{\nabla}^\prime\cdot\bm{u}^\prime \right)\right\rangle=0.
\end{align*}

Again, similar to Eq.~\eqref{E_phi3}, one can also obtain an evolution equation for $\bm{q}^\prime$. Combining $\bm{q}\cdot\partial_t\bm{q}^\prime$ and $\bm{q}^\prime\cdot\partial_t\bm{q}$, we get
\begin{align}
\partial_t &\left\langle\bm{q}\cdot \bm{q}^\prime \right\rangle \nonumber\\
= &\left\langle  -\bm{q}^\prime\cdot\bm{\nabla}(S u_z+ \bm{u}\cdot\bm{q})- \bm{q}\cdot\bm{\nabla}^\prime( Su^\prime_z+ \bm{u}^\prime\cdot\bm{q}^\prime) \right\rangle \nonumber\\
&+ D_q + F_q \label{RE_q1} \\
=  &- \left\langle\bm{\nabla} \cdot[S u_z  \bm{q}^\prime+ (\bm{u}\cdot\bm{q}) \bm{q}^\prime]+\bm{\nabla}^\prime\cdot[S u^\prime_z  \bm{q}+(\bm{u}^\prime\cdot\bm{q}^\prime) \bm{q}] \right\rangle\nonumber\\
&+  D_q + F_q \label{RE_q2}\\
=&-\bm{\nabla}_{\bm{r}}\cdot\left\langle Su_z^\prime\bm{q}-Su_z\bm{q}^\prime+(\bm{u}^\prime\cdot\bm{q}^\prime) \bm{q} -(\bm{u}\cdot\bm{q}) \bm{q}^\prime\right\rangle\nonumber\\
&+D_q+F_q,
\label{RE_q}
\end{align}
where $D_q = \langle\bm{q}^\prime\cdot\mathcal{M}\nabla^2\left(\bm{\nabla}\mu\right) +\bm{q}\cdot\mathcal{M}{\nabla^\prime}^2\left(\bm{\nabla}^\prime\mu^\prime\right)\rangle$ and $F_q = \langle\bm{q}^\prime\cdot\bm{g}+\bm{q}\cdot\bm{g}^\prime\rangle$. In the following, we derive the exact relation in two ways \citep{banerjee2017exact}: 

 \subsection{In terms of two-point increments} \label{exact1}

\noindent Adding Eqs. \eqref{RE_u} and \eqref{RE_q}, we get
\begin{align}
 \partial_t \mathcal{R} =&-\frac{1}{2} \bm{\nabla}_{\bm{r}}\cdot \left\langle( \bm{u}\cdot \bm{u}^\prime)\delta\bm{u}+\xi[( \bm{u}\cdot\bm{q}^\prime)\bm{q}^\prime
 -(\bm{u}^\prime\cdot \bm{q})\bm{q} \nonumber\right.\\  
 & +\left.(\bm{u}^\prime\cdot\bm{q}^\prime) \bm{q}-(\bm{u}\cdot\bm{q})\bm{q}^\prime]\right\rangle + D + F, \label{RE_F1}
 \end{align}
where $\mathcal{R} = \left(\mathcal{R}_E + \mathcal{R}^{\prime}_E\right)/2$, $D = \left(D_u + D_q\right)/2$ and $F =\left( F_u + F_q\right)/2$. Furthermore, under statistical homogeneity, we obtain
{\small
\begin{equation}
    \bm{\nabla}_{\bm{r}}  \cdot \left\langle(\bm{u}^\prime \cdot \bm{q})\bm{q}^\prime\right\rangle 
= -\left\langle \bm{q}^\prime\cdot (\bm{u}^\prime\cdot\bm{\nabla})  \bm{q} \right \rangle 
= \bm{\nabla_r}\cdot\langle \bm{u}^\prime (\bm{q}\cdot\bm{q}^\prime) \rangle, \label{Identity1}
\end{equation}}
where we use ${\bm \nabla} \times {\bf q}= {\bm 0}$ along with the identity $\bm{\nabla}(\bm{A}\cdot\bm{B})= (\bm{A}\cdot\bm{\nabla})\bm{B}+(\bm{B}\cdot\bm{\nabla})\bm{A} + \bm{A}\times(\bm{\nabla}\times\bm{B})+\bm{B}\times(\bm{\nabla}\times\bm{A})$. Now, we consider a statistically stationary state where the $l.h.s$ of the Eq.~\eqref{RE_F1} vanishes. In the limit of infinite Reynolds number, within the inertial range, the dissipative effects can also be neglected and the corresponding energy flux rate $\varepsilon$ can be associated with the total energy injection rate as $F= \varepsilon$. Finally, using Eq.~\eqref{Identity1} and following some straightforward algebra, we can express the two-point correlations of Eq.~\eqref{RE_F1} in terms of the two-point increments whence the final exact relation can be obtained as
{\small
\begin{equation}
\bm{\nabla}_{\bm{r}}  \cdot \left \langle \left[ (\delta \bm{u})^2   - \xi (\delta \bm{q})^2 \right] \delta \bm{u}   +2\xi (\delta \bm{u}\cdot \delta \bm{q}) \delta \bm{q}\right\rangle = -4\varepsilon. \label{RE_F2}
\end{equation}}
Eq.~\eqref{RE_F2} is the main result of our paper. 
As mentioned in the introduction, here the exact relation is cast in a differential form with 
\begin{align}
    \bm{\mathcal{F}}\equiv \left \langle \left[  (\delta \bm{u})^2   - \xi (\delta \bm{q})^2 \right] \delta \bm{u}   +2\xi (\delta \bm{u}\cdot \delta \bm{q}) \delta \bm{q}\right\rangle.
\end{align}
It expresses the inertial range energy flux rate $\varepsilon$ purely in terms of two-point increments of the field variables of BFT. 
Therefore, $\varepsilon$ remains unchanged if ${\bf q}$ is replaced by $\bm \nabla \phi$ in Eq.~\eqref{RE_F2}. For all simple and active binary fluids with extensile stress, $\xi$ is positive and so the form of conserved energy remains same as Eq.~\eqref{total_energy}
\begin{align}
    E= \int \frac{1}{2} \left(u^2 + |\xi| q^2\right) d\tau, \label{energy2}
\end{align}
and the corresponding exact law is given by
{\small
\begin{equation}
  \bm{\nabla}_{\bm{r}}  \cdot \left \langle \left[  (\delta \bm{u}) ^2   - |\xi| (\delta \bm{q})^2\right] \delta \bm{u}   + 2|\xi| (\delta \bm{u}\cdot \delta \bm{q}) \delta \bm{q}\right \rangle = -4 \varepsilon.  \label{RE_F3a}
\end{equation}
}
However, active fluids with large contractile stress ($\eta<0$ and $|\eta|>|\kappa|$) have $\xi$ to be negative. In that case, the total energy becomes
\begin{align}
    E= \int \frac{1}{2} \left(u^2 - |\xi| q^2\right) d\tau, \label{energy1}
\end{align}
and the corresponding exact relation becomes
{\small
\begin{equation}
  \bm{\nabla}_{\bm{r}}  \cdot \left \langle \left[  (\delta \bm{u}) ^2   + |\xi| (\delta \bm{q})^2\right] \delta \bm{u}   - 2|\xi| (\delta \bm{u}\cdot \delta \bm{q}) \delta \bm{q}\right \rangle = -4 \varepsilon.  \label{RE_F3}
\end{equation}
}
Surprisingly, Eq.~\eqref{RE_F3} looks very much similar to the exact relation for energy transfer in incompressible MHD turbulence if one replaces the field $\sqrt{\vert \xi \vert} {\bf q}$ by the local magnetic field in Alfv\'en units \citep{politano1998dynamical}. This can be a bit misleading as for MHD, the same substitution in Eq.~\eqref{energy1} would actually correspond to the residual energy which is not an inviscid invariant of incompressible MHD.  
\paragraph*{\bf Passive scalar flow:}
In the limit where the activity coefficient $\xi$ tends to zero, the Eqs.~\eqref{RE_F2} and \eqref{RE_F3} simply reduce to 
\begin{equation}
  \bm{\nabla}_{\bm{r}}  \cdot \left \langle  (\delta \bm{u}) ^2  \delta \bm{u}  \right \rangle = -4 \varepsilon.  \label{RE_F3b}
\end{equation}
which represents the inertial range energy transfer in a passive scalar flow and is identical to that in incompressible HD turbulence. In fact by putting $\xi=0$, one turns off the feedback force in the momentum evolution equation due to the scalar field $\phi$ thus leading to the individual conservation of kinetic energy similar to the incompressible HD case. 

\paragraph*{\bf Weakly correlated fluctuations:} 
In case $\delta\bm{u}$ and $\delta\bm{q}$ are weakly correlated, we have $ \left\langle(\delta \bm{u}\cdot \delta \bm{q}) \delta \bm{q}\right\rangle \simeq 0$ and the exact relation \eqref{RE_F2} practically becomes 
\begin{equation}
\bm{\nabla}_{\bm{r}}  \cdot \left \langle \left[ (\delta \bm{u})^2   - \xi (\delta \bm{q})^2 \right] \delta \bm{u}   \right\rangle = -4\varepsilon. \label{RE_F2a} 
\end{equation}
Further if we assume $|\delta\bm{u}|\sim |\delta\bm{q}|$, then we can encounter two interesting situations. First, when $\xi>0$ but very small ($e.g$ simple binary fluids \citep{Ruiz1981turbulence}), then the flux $\bm{\mathcal{F}}\simeq \left \langle (\delta \bm{u})^2 \delta \bm{u} \right\rangle$ and hence corresponds to the usual Kolmogorov case. However, if $\xi$ becomes sufficiently large  $e.g.$ for active binary fluids with extensile stress,  $\bm{\mathcal{F}}\simeq \left \langle- \xi (\delta \bm{q})^2 \delta \bm{u} \right\rangle$ thereby leading to the possibility of an inverse cascade of energy. 
In contrast for active binary fluids with contractile stress where $\xi<0$, we have $\bm{\mathcal{F}}\simeq \left \langle \left[ (\delta \bm{u})^2 + |\xi| (\delta \bm{q})^2 \right] \delta \bm{u}   \right\rangle$, a direct cascade of energy cascade is always expected. {\color{black}Note that, unlike two dimensional hydrodynamics, here we are
not talking about simultaneous forward and inverse cascades of two invariants. In the present case, the energy cascade is forward or inverse depending upon the activity parameter $\xi$. For a system with given $\xi$, the cascade direction is therefore automatically determined.}

However, such type of speculations can be non-trivial if the correlation between $\delta\bm{u}$ and $\delta\bm{q}$ is not negligible. The direction of cascade then depends on the mutual competition of the various terms in the flux and can be explored numerically. Such  investigation certainly demands a separate study and is beyond the scope of the present paper which presents a systematic analytical approach for an exact calculation of the inertial range energy transfer rate in homogeneous BFT.   
%However, the cons$\frac{1}{2}\langle \phi^2 \rangle $ is always conserved and the exact relation corresponding to this conserved quantity will be exactly same as derived by Yaglom \citep{Yaglom1949} for passive scalars
%\begin{equation}
 % \bm{\nabla}\bm{r} \cdot \langle  \delta \phi ^2  \delta \bm{u} \rangle = -4 \varepsilon. 
%\end{equation}

\subsection{In terms of two-point correlators}
In order to obtain the exact relations in terms of two-point correlators we start from Eq.~\eqref{RE_u1} and~\eqref{RE_q1}. Splitting the two-point energy correlator $\mathcal{R}$ in the kinetic and active energy correlators $\mathcal{R}_u$ and $\mathcal{R}_q$ respectively, we can write the evolution equations as follows:
\begin{align}
\partial_t \mathcal{R}(\bm{r},t)&= \mathcal{T}_u+\mathcal{T}_q+D+ F,\\
\partial_t \mathcal{R}_u(\bm{r},t)&= \mathcal{T}_u+\mathcal{\chi}_{qu}+D_u+ F_u,\\
\partial_t \mathcal{R}_q(\bm{r},t)&= \mathcal{T}_q-\mathcal{\chi}_{qu}+D_q+ F_q,
\end{align}
where 
\begin{align}
\mathcal{T}_u(\bm{r},t)=&\frac{1}{2}\left\langle -\bm{u}^\prime \cdot(\bm{u}\cdot\bm{\nabla}) \bm{u} -\bm{u}\cdot(\bm{u}^\prime\cdot\bm{\nabla}^\prime) \bm{u}^\prime\right.\nonumber\\
-&\left.\xi(\bm{u}^\prime\cdot\bm{q}) 
(\bm{\nabla}\cdot\bm{q})
 -\xi(\bm{u}\cdot\bm{q}^\prime ) (\bm{\nabla}^\prime \cdot\bm{q}^\prime) \right\rangle, \\
 \mathcal{\chi}_{uq}(\bm{r},t)=& \frac{1}{2}\left\langle -\xi Su_z^\prime (\bm{\nabla}\cdot\bm{q}) -\xi Su_z (\bm{\nabla}^\prime\cdot\bm{q}^\prime) \right\rangle,\\
\mathcal{T}_q(\bm{r},t)=&\frac{1}{2}\left\langle -\bm{q}^\prime\cdot\bm{\nabla}( \bm{u}\cdot\bm{q})- \bm{q}\cdot\bm{\nabla}^\prime( \bm{u}^\prime\cdot\bm{q}^\prime)\right\rangle,
\end{align}
with $\mathcal{T}_{u}(\bm{r},t)$ and $\mathcal{T}_{q}(\bm{r},t)$ being the scale-to-scale kinetic and active energy transfer terms and $\mathcal{\chi}_{qu}(\bm{r},t)$ being the active to kinetic energy conversion term. Similar to Eqs.~\eqref{RE_F2} and \eqref{RE_F3}, it is evident to see that the mean gradient field ($S\hat{z}$), which appears only in the conversion terms, can not affect the scale-to-scale energy transfer rate $\varepsilon$. This is similar to incompressible MHD turbulence where the mean magnetic field can not alter the turbulent energy transfer. Reasoning as in the previous subsection here the final expression of the inertial range exact relation can be written as 
\begin{equation}
    \mathcal{T}_{u} + \mathcal{T}_{q}= - \varepsilon. \label{correlator}
\end{equation}
Eq.~\eqref{correlator} is another important result of this paper. This form is particularly useful for calculating $\varepsilon$ using spectral method \citep{banerjee2017exact}. 

\subsection{In terms of new variables \boldmath\texorpdfstring{\color{black}{$\Upsilon^{\pm}$}}{}}\label{upsilon variables}
Instead of $\bm{u}$ and $\bm{b}$, the equations for incompressible MHD can also be written in terms of Els\"{a}sser variables $\bm{z}^{\pm} = \bm{u}\pm\bm{b}$ \citep{Elsasser1950hydrodynamic}. In a similar way, the basic equations for BFT can also be written in terms of ‘upsilon’ variables $\bm{\Upsilon}^{\pm} = \bm{u}\pm i\bm{Q}$, where $\bm{Q} = \sqrt{\xi}\bm{\nabla}\phi$. This can be obtained by writing the Eqs.~\eqref{E_momentum3} and~\eqref{E_phi2} in the following form
\begin{widetext}
\begin{align}
  \partial_t \bm{u} = &-(\bm{u}\cdot\bm{\nabla})\bm{u} + (\bm{Q}\times\bm{\nabla})\times\bm{Q}- \bm{\nabla} P_T + \nu \nabla^2 \bm{u} + \bm{f}, \label{Evolution_u_symmetric}\\
    \partial_t\bm{Q} =& -(\bm{u}\cdot\bm{\nabla})\bm{Q} - (\bm{Q}\times\bm{\nabla})\times\bm{u}  + D\nabla^2 \bm{Q} + \bm{g}, \label{Evolution_phi_symmetric}
\end{align}
\end{widetext} where $P_T = P^* + Q^2$ and we use the following vector-calculus identity $(\bm{A}\times\bm{\nabla})\times\bm{B}=\bm{A}\times(\bm{\nabla}\times\bm{B})+ (\bm{A}\cdot\bm{\nabla})\bm{B} - \bm{A}(\bm{\nabla}\cdot\bm{B})$. Note that here we are considering the chemical potential $\sqrt{\xi}\bm{\nabla}\mu\sim \bm{Q}$ which is practically true for phase-mixed or nearly phase separating binary fluid.
Finally combining the Eqs. \eqref{Evolution_u_symmetric} and  \eqref{Evolution_phi_symmetric}, one can write the basic equations of BFT in terms of the $\bm{\Upsilon^{\pm}}$ variables as:
\begin{widetext}
\begin{align}
    \partial_t \bm{\Upsilon^{+}} =& \frac{1}{2} \left[ -\{(\bm{\Upsilon^{+}}+\bm{\Upsilon^{-}})\cdot \bm{\nabla}\} \Upsilon^{+}- \{(\bm{\Upsilon^{+}}-\bm{\Upsilon^{-}})\times \bm{\nabla}\}\times\bm{\Upsilon^{+}}
    +\nabla^2(\nu_{+} \bm{\Upsilon^{+}}+\nu_{-} \bm{\Upsilon^{-}}) \right]-\bm{\nabla}P_T + \bm{f}_{+}, \label{Evolution_Upsilon_plus}\\
     \partial_t \bm{\Upsilon^{-}} =& \frac{1}{2} \left[ -\{(\bm{\Upsilon^{+}}+\bm{\Upsilon^{-}})\cdot \bm{\nabla}\} \Upsilon^{-}
   + \{(\bm{\Upsilon^{+}}-\bm{\Upsilon^{-}})\times \bm{\nabla}\}\times\bm{\Upsilon^{-}} +\nabla^2(\nu_{-} \bm{\Upsilon^{+}}+\nu_{+} \bm{\Upsilon^{-}})\right]-\bm{\nabla}P_T + \bm{f}_{-}, \label{Evolution_Upsilon_minus}
\end{align}
\end{widetext}
where $\nu_{\pm} = (\nu \pm D)/2$ and $\bm{f}_{\pm} = \bm{f} \pm i\bm{g}$. The above equations show that, unlike the {\color{black}Els\"{a}sser} variables in incompressible MHD, $\bm{\Upsilon}^{\pm}$ undergo both cross and self-deformation through two types of non-linear interactions. In addition to the usual advective non-linear interactions $(\bm{u}\cdot\bm{\nabla})\bm{\Upsilon}^{\pm}$, here we also have other non-linear terms proportional to $(\bm{Q}\times\bm{\nabla})\times\bm{\Upsilon}^{\pm}$. For a given direction of $\bm{Q}$ (say \,$\hat{z}$), $(\bm{Q}\times\bm{\nabla})\times\bm{\Upsilon}^{\pm}$ calculate the variation of $\bm{\Upsilon}^{\pm}$ in a plane perpendicular to that direction. In contrast to the IK phenomenology \citep{Iroshnikov1964turbulence,Kraichnan1965Inertial} for incompressible MHD turbulence, the co-existence of two aforesaid non-linear interactions and their possible entanglement does not provide a simple phenomenological image for BFT. However, in Sec. \ref{Discussion}, we predict a power law dependence of the conserved energy analogous to IK  phenomenology based on the relative importance of various time scales corresponding to different interactions.
From definition, one can immediately show \begin{equation}
    \bm{\nabla}\cdot\bm{\Upsilon^{\pm}}= \pm i \bm{\nabla}\cdot\bm{Q}, \;  \; \bm{\nabla}\times \bm{\Upsilon^{\pm}} = \bm{\nabla} \times \bm{u}.
\end{equation}
This is also in contrast with the {\color{black}Els\"{a}sser} variables which are divergence-free similar to the $\bm{u}$ and $\bm{b}$ fields of incompressible MHD. 

Previously, we showed that the turbulent energy is an inviscid invariant of a binary fluid flow. By the same method, one can also show the total energy
\begin{equation}
    E = \int \frac{1}{2}(u^2+ Q^2)d \tau
\end{equation}
is (as expected) an inviscid invariant. In terms of the upsilon variables, the total energy becomes 
$ E= \int\frac{1}{2}(\bm{\Upsilon}^{+}\cdot\bm{\Upsilon}^{-})d\tau$.
Furthermore, the exact relation derived in Eq.~(\ref{RE_F2}) can be written as  
{\small
\begin{align}
      -4 \varepsilon =& \bm{\nabla}_{\bm{r}}  \cdot \left \langle \left[  (\delta \bm{u}) ^2   -  (\delta \bm{Q})^2\right] \delta \bm{u}   + 2(\delta \bm{u}\cdot \delta \bm{Q}) \delta \bm{Q} \right \rangle \nonumber\\
      =& \frac{1}{8}\bm{\nabla}_{\bm{r}}  \cdot \left \langle 2\left[  (\delta \bm{\Upsilon}^{+})^2 + (\delta \bm{\Upsilon}^{-})^2 \right]  (\delta\bm{\Upsilon}^{+}+\delta\bm{\Upsilon}^{-})\right.\nonumber\\
      &-\left. 2 [(\delta \bm{\Upsilon}^{+})^2 - (\delta \bm{\Upsilon}^{-})^2 ]  (\delta\bm{\Upsilon}^{+}-\delta\bm{\Upsilon}^{-}) \right \rangle \nonumber\\
     =& \frac{1}{2} \bm{\nabla}_{\bm{r}} \cdot \left \langle (\delta \bm{\Upsilon}^{-})^2 \delta \bm{\Upsilon}^{+} + (\delta \bm{\Upsilon}^{+})^2 \delta \bm{\Upsilon}^{-}  \right \rangle. \label{RE_Upsilon_Positive}
\end{align}} 
Note that, the same exact relation can also be derived directly from the Eqs.~\eqref{Evolution_Upsilon_plus} and \eqref{Evolution_Upsilon_minus}. Interestingly Eq.~\eqref{RE_Upsilon_Positive} looks very similar to the exact relation of  energy transfer in incompressible MHD turbulence when expressed in terms of the Els\"{a}sser variables \citep{banerjee2013exact,banerjee2014compressible}. Here one has to remember that the Els\"asser fields ${\bm z}^{\pm}$ are always real. This is not true for the upsilon variables. For $\xi>0$, one can write  $\bm{\Upsilon}^{\pm} =\bm{u}\pm i \sqrt{|\xi|}\bm{\nabla}\phi $ thus yielding complex upsilon fields. However, for $\xi<0$, we have $\bm{\Upsilon}^{\pm} = \bm{u}\mp \sqrt{|\xi|}\bm{\nabla}\phi $ and hence the upsilon variables become real in that case. Irrespective of whether the upsilon fields are real or complex, it is straightforward to verify that the flux term in the \textit{r.h.s.} of Eq.~\eqref{RE_Upsilon_Positive} is, as expected, always real.

\section{Discussion}\label{Discussion}
{\color{black}In this paper we derived several exact relations} for fully developed, three dimensional, homogeneous binary fluid turbulence. Using this relation, we can calculate the scale-to-scale transfer rate of total energy (kinetic plus active energy) within the so-called inertial range. Previously, it has been argued  \citep{Ruiz1981turbulence} that simple binary fluids and incompressible MHD are structurally similar on account of the equations of dynamics and linear wave modes. Drawing analogy with IK phenomenology, they also predicted a $-3/2$ power law for turbulent energy spectra. However, for fully developed turbulence, our paper shows that the generic form of the exact relation derived in Eq.~\eqref{RE_F2} differs from that of incompressible MHD turbulence \citep{politano1998dynamical}. In particular, the flux $\bm{\mathcal{F}}$ in Eq.~\eqref{RE_F2} has two sign reversals in comparison with that of Eq.~\eqref{Incompressible_MHD}. 
%corresponding fluxes $\bm{\mathcal{F}}$ have opposite sign for active energy terms $\left\langle- \xi(\delta \bm{q})^2 \delta \bm{u}\right\rangle$ and $ 2\xi (\delta \bm{u}\cdot\delta \bm{q})$ in Eqs.~\eqref{RE_F3a} and ~\eqref{Incompressible_MHD} respectively.
 Interestingly, active binary fluids with contractile stress ($\xi<0$) is found to be algebraically identical to the exact relation of incompressible MHD turbulence if we replace the magnetic fields in Eq.~\eqref{Incompressible_MHD} with $\sqrt{|\xi|}\bm{q}$. Nevertheless, this two systems are categorically different with respect to the linear stability analysis. Under weak perturbations, an active binary fluid with contractile stress leads to linear instability whereas an incompressible MHD fluid responds to weak perturbations in terms of Alfv\'en waves. 
From Eq.~\eqref{RE_F2}, we retrieved the exact relation for passive scalar turbulence in the limit of vanishing activity parameter. In addition, we have predicted a possible inverse cascade of energy in 3D active binary fluid turbulence with extensile stress when the correlation between the $\bm{u}$ and the $\bm{q}$ fields are sufficiently weak.
Inspired by the Els\"asser variables, here we introduced the `upsilon' variables and wrote the dynamical equations for binary fluids in a more symmetric form by the introduction of upsilon variables. Finally, from Eq.~\eqref{RE_F2}, we also wrote the exact relation in terms of the upsilon variables. Interestingly, this exact relation looks exactly similar to that of incompressible MHD turbulence when expressed in terms of the Els\"asser variables.
However unlike incompressible MHD, here the cross-helicity $\int (\bm{u}\cdot\bm{q}) d \tau$ is not an inviscid invariant. A helicity cascade is therefore not guaranteed \citep{frisch1995turbulence} and hence the derivation of the corresponding exact relation is not useful in BFT. 

Based on our previous analysis, here we propose a plausible phenomenology and predict a power law for {\color{black}the turbulent energy spectrum in simple and active binary fluids with extensile stress}. Writing the Eqs.~\eqref{Evolution_Upsilon_plus} and \eqref{Evolution_Upsilon_minus} in a compact form, we obtain
\begin{align}
    \partial_t \bm{\Upsilon^{\pm}} = & -(\bm{u}\cdot \bm{\nabla}) \Upsilon^{\pm}\mp i(\bm{Q}\times \bm{\nabla})\times\bm{\Upsilon^{\pm}}
    -\bm{\nabla}P_T \nonumber\\
    &+\nabla^2(\nu_{\pm} \bm{\Upsilon^{+}}+\nu_{\mp} \bm{\Upsilon^{-}}) + \bm{f}_{\pm},\label{Evolution_Upsilon_pm}
\end{align}
In the above equations, possible nonlinear interactions can be obtained from the terms $(\bm{u}\cdot \bm{\nabla}) \Upsilon^{\pm}$ and $(\bm{Q}\times \bm{\nabla})\times\bm{\Upsilon^{\pm}}$. Whereas $(\bm{u}\cdot \bm{\nabla}) \Upsilon^{\pm}$ represent the advection of $\bm{\Upsilon}^{\pm}$ by the velocity field, the terms $(\bm{Q}\times \bm{\nabla})\times\bm{\Upsilon^{\pm}}$ represent the variation of $\bm{\Upsilon}^{\pm}$ in a plane perpendicular to $\bm{Q}$. Expressing $\bm{Q}$ as a sum of the mean field $\bm{Q}_0$ and the fluctuation $\widetilde{\bm{Q}}$ in  Eq.~\eqref{Evolution_Upsilon_pm}, we get three types of interactions. While we associate two kinds of non-linear time scales $\tau^{u}_{\ell}$ and $\tau^{Q}_{\ell}$ corresponding to the terms $(\bm{u}\cdot \bm{\nabla}) 
\widetilde{\bm{\Upsilon}}^{\pm}$ and $(\widetilde{\bm{Q}}\times \bm{\nabla})\times \widetilde{\bm{\Upsilon}}^{\pm}$ respectively,
one linear time-scale $\tau_{0\ell}$ corresponds to the term $(\bm{Q}_0\times \bm{\nabla})\times \widetilde{\bm{\Upsilon}}^{\pm}$, with $\widetilde{\bm{a}}$ representing the fluctuating part of the vector $\bm{a}$. {\color{black} This linear time can be associated with the concentration waves as defined in \citep{Ruiz1981turbulence} and the corresponding dispersion relation is given by $\omega(k) = \pm \sqrt{\xi} |\bm{k}\times \bm{Q}_0|$}. If the two non-linear interactions are assumed to be independent, the effective distortion time-scale $\tau_{\ell}$ becomes
\begin{align}
1/\tau_{\ell}=1/\tau^{u}_{\ell} + 1/\tau^{Q}_{\ell},
\end{align}
where $\ell (\equiv |\bm{r}|)$ is the characteristic size of the eddies, $u_{\ell}\sim |\delta \bm{u}|$, $Q_{\ell}\sim |\delta \bm{Q}|$, $\tau^{u}_{\ell}\sim \ell/u_{\ell}$ and $\tau^{Q}_{\ell}\sim \ell/\widetilde{Q}_{\ell}$. Analogous to the Alfv\'en time scale in incompressible MHD, here we can also define the linear time scale $\tau_{0\ell}\sim {\ell}/Q_{0}$.
In the presence of strong $\bm{Q}_0$, we have $\tau_{0\ell}<<\tau_{\ell}$. In that case, the energy transfer time-scale $\tau^{tr}_{\ell} \sim \tau_{\ell}^2/{\color{black}\tau_{0\ell}}$ and the energy flux rate $\varepsilon$ would scale as \citep{Biskamp2003magnetohydrodynaics,Iroshnikov1964turbulence,Kraichnan1965Inertial}
\begin{align}
    \varepsilon \sim ( \widetilde{\Upsilon}^{+}_{\ell}  \widetilde{\Upsilon}^{-}_{\ell} )/ \tau^{tr}_{\ell}.\label{energy_flux}
\end{align}
Under the assumption of weak correlations between $u_{\ell}$ and $Q_{\ell}$, we have  $\widetilde{\Upsilon}^{+}_{\ell}\sim \widetilde{\Upsilon}^{-}_{\ell} \sim \widetilde{\Upsilon}_{\ell} \sim u_{\ell}$ and hence $\tau_{\ell} \sim \tau^{u}_{\ell}\sim {\ell}/\widetilde{\Upsilon}_{\ell}$. Combining all, finally we can write 
\begin{equation}
     \varepsilon \sim  {\widetilde{\Upsilon}_{\ell}}^4/ {\ell} Q_{0} \Rightarrow \widetilde{\Upsilon}_{\ell} \sim (\varepsilon Q_{0})^{1/4} {\ell}^{1/4}.
\end{equation}
Again, by definition of energy spectrum $E(k)$
\begin{equation}
   \widetilde{\Upsilon}^2_{\ell} \sim {\color{black}E(k)k} \Rightarrow
    E(k) \sim (\varepsilon Q_{0})^{1/2}k^{-3/2}, 
\end{equation}
where $k$ is the wavenumber corresponding to $\ell$. This is in agreement with the predictions of \citep{Ruiz1981turbulence}, which was done for simple binary fluids only. 

Similar types of studies can be generalized to the compressible binary fluids as well as to the mixtures of more than two fluids.

\section{Acknowledgement}
The authors acknowledge Jayanta K. Bhattacharjee for useful discussions. NP acknowledges the grant under PMRF scheme (Government of India). SB acknowledges the DST-Inspire faculty research grant (DST/PHY/2017514).

\bibliographystyle{apsrev4-2}
\bibliography{Driven}% Produces the bibliography via BibTeX.

\end{document}